\documentclass[fleqn,usenatbib]{mnras}
\usepackage{newtxtext,newtxmath}
\usepackage[T1]{fontenc}
\usepackage{gensymb}

\DeclareRobustCommand{\VAN}[3]{#2}
\let\VANthebibliography\thebibliography
\def\thebibliography{\DeclareRobustCommand{\VAN}[3]{##3}\VANthebibliography}

\usepackage{CJKutf8}
\usepackage{booktabs}
\usepackage{xcolor, soul}
\usepackage{amsmath, bm}
\usepackage{graphicx}

\title[Weak lensing with Scattering Transform]
{Weak lensing scattering transform: dark energy and neutrino mass sensitivity}

\author[Cheng \& M\'enard]{
Sihao Cheng (程思浩)$^{1}$\thanks{E-mail: s.cheng@jhu.edu} \&
Brice M\'enard$^{1}$
\\
$^{1}$Department of Physics and Astronomy, The Johns Hopkins University, 3400 N Charles Street, Baltimore, MD 21218, USA\\
}

\date{\today} 
\pubyear{2021}

\begin{document}
\label{firstpage}
\pagerange{\pageref{firstpage}--\pageref{lastpage}}
\begin{CJK}{UTF8}{gkai} 
\maketitle
\end{CJK}

\begin{abstract}
As weak lensing surveys become deeper, they reveal more non-Gaussian aspects of the convergence field which can only be extracted using statistics beyond the power spectrum. In \cite{Cheng_2020}, we showed that the scattering transform, a novel statistic borrowing mathematical concepts from convolutional neural networks, is a powerful tool for cosmological parameter estimation in the non-Gaussian regime. Here, we extend that analysis to explore its sensitivity to dark energy and neutrino mass parameters with weak lensing surveys. We first use image synthesis to show visually that, compared to the power spectrum and bispectrum, the scattering transform provides a better statistical vocabulary to characterize the perceptual properties of lensing mass maps. 
We then show that it is also better suited for parameter inference:
(i) it provides higher sensitivity in the noiseless regime, and 
(ii) at the noise level of Rubin-like surveys, though the constraints are not significantly tighter than those of the bispectrum, the scattering coefficients have a more Gaussian sampling distribution, which is an important property for likelihood parametrization and accurate cosmological inference. We argue that the scattering coefficients are preferred statistics considering both constraining power and likelihood properties.
\end{abstract}


\section{Introduction}
\label{sec:intro}

Statistical properties of matter density distribution in our Universe carry information about its components and evolutionary history.
In the late-time universe probed by weak gravitational lensing distortion, or cosmic shear \citep[see][for reviews]{Bartelmann_2001, Kilbinger_2015}, the non-linear growth of structure imprints non-Gaussianity at small scales. As a result, cosmological information starts to elude Gaussian statistics such as the power spectrum and correlation function \citep{Rimes_2005, Rimes_2006, Neyrinck_2006, Neyrinck_2007}. 
To extract more information, some non-Gaussian statistics, such as the bispectrum \citep{Fu_2014}, peak counts \citep{LPH15, LPL15, Kacprzak_2016, Martinet_2018, Shan_2018}, and Minkowski functionals \citep{Petri_2015}, have been applied to existing weak lensing data. Deeper ongoing and future surveys like the Hyper Suprime-Cam survey \citep[HSC, ][]{Aihara_2018a,Aihara_2018b}, Rubin Observatory survey \citep[formerly LSST,][]{LSST_2009}, \textit{Euclid} \citep{Laureijs_2011}, and Roman space telescope survey \citep[formerly WFIRST,][]{Spergel_2015} have access to even smaller scales, making it increasingly important to employ proper statistics that can capture the non-Gaussian information in lensing maps \citep[see, e.g.,][and reference therein]{Martinet_2021}.

In \cite{Cheng_2020}, we proposed to use a non-Gaussian statistical tool that is novel to cosmology: the scattering transform\footnote{To our best knowledge, the scattering transform has nothing to do with the scattering process in particle physics.}. We demonstrated its power by applying it to simulated weak lensing maps to constrain the $\Omega_m$ and $\sigma_8$ cosmological parameters. In this companion paper, we extend our analysis to other cosmological dependencies: the dark energy equation of state $w_0$, $w_a$ and the neutrino mass sum $M_\nu$. Constraining these parameters is one of the key goals of upcoming weak lensing surveys and they are often used in the figure of merits to compare different survey strategies.

The scattering transform, introduced by \citep{Mallat_2012} in the signal processing literature, uses concepts found in convolutional neural networks, but it needs no training: similar to traditional statistical estimators, it generates a set of summary statistics, which efficiently characterize complex non-Gaussianity at various scales. To visualize the power of scattering coefficients, we will first show that one can use them to generate random images with textures very similar to real lensing maps, to a level not achievable by traditional moment-based statistics such as the power spectrum and bispectrum. Then, we will present the constraining power of scattering coefficients for dark energy parameters and neutrino mass through a forecast using simulated lensing maps.

\section{Scattering transform}
\label{sec:ST}

\begin{figure}
    \centering
    \includegraphics[width=\columnwidth]{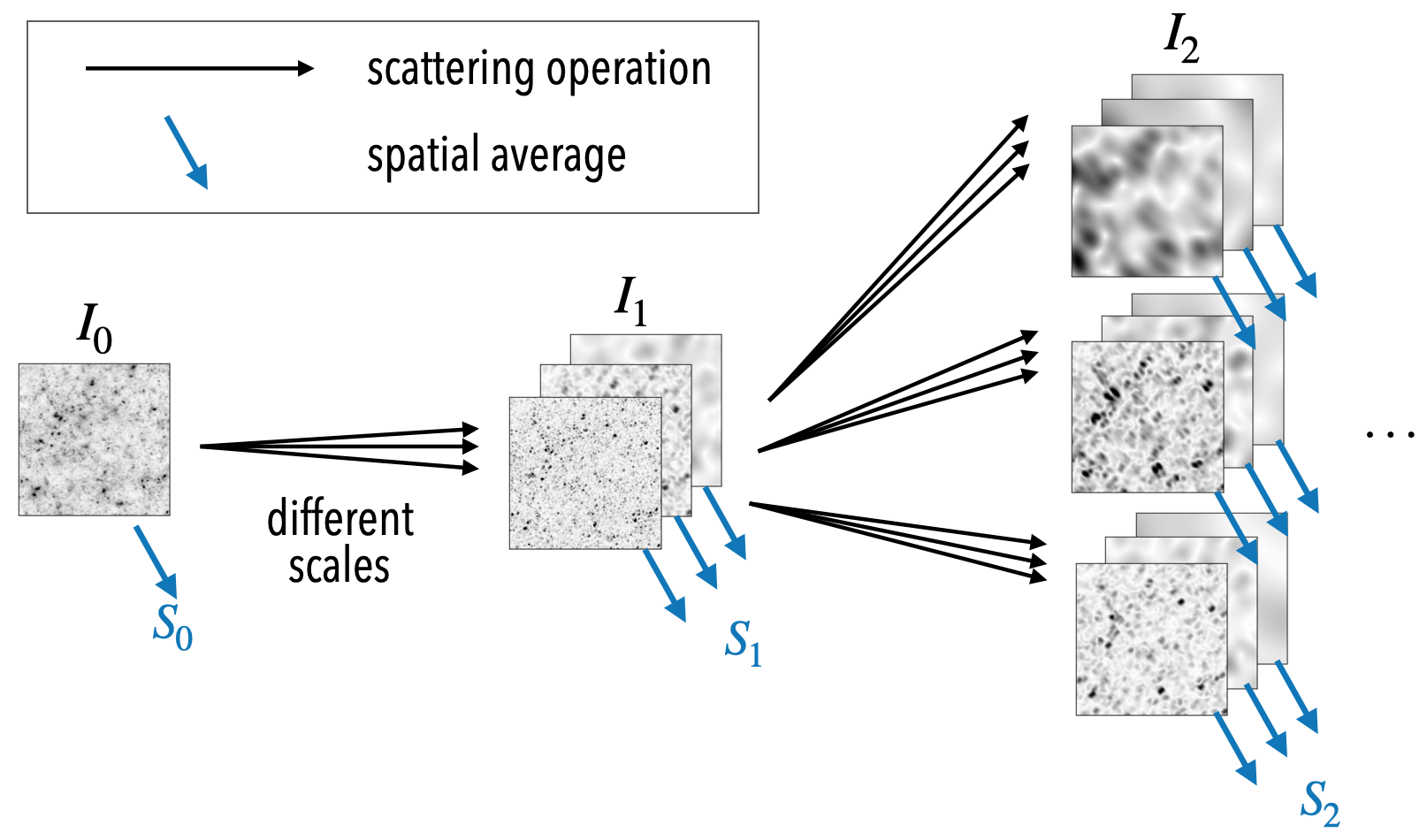}
    \caption{Structure of the scattering transform. From the input field, a tree of intermediate fields are generated by iterating the scattering operation, then the spatial averages of these fields, are taken as the translation-invariant scattering coefficients. The scattering operation is composed of a wavelet convolution and a pixelwise modulus.}
    \label{fig:ST_tree}
\end{figure}

\begin{figure*}
\begin{center}
    \includegraphics[width=0.97\textwidth]{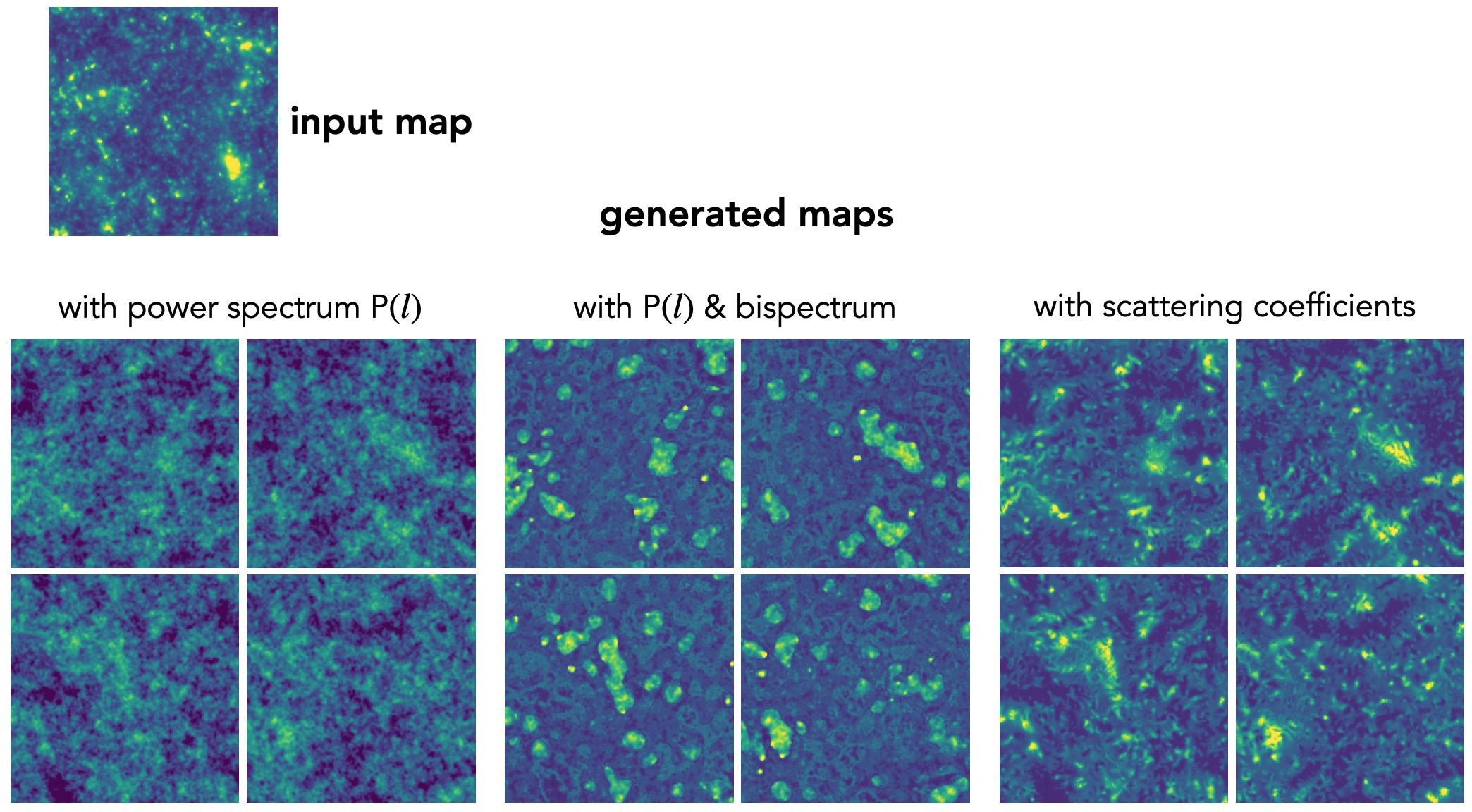}
    \vspace{-.2cm}
    \caption{Images generated with selected sets of summary statistics, including 
    the power spectrum, bispectrum, and scattering coefficients ($S_1$, $S_2$), showing that the scattering coefficients characterize the field better. As the image generation is random, one should compare textures, instead of the exact positions of features, between the input and generated maps.}
    \label{fig:synthesis}
\end{center}
\end{figure*}

The scattering transform \citep{Mallat_2012, Bruna_2013} was originally developed in the context of signal processing in computer vision \citep[see, e.g.,][]{Sifre_2013, AndenMallat_2014, Bruna_2015}. In astrophysics, it has been used for parameter inference in cosmology \citep{Cheng_2020} and analysis of interstellar medium \citep{Allys_2019, Saydjari_2020, Blancard_2020}.

The scattering transform shares some properties with both convolutional neural networks (CNNs) and $N$-point correlation functions in desirable ways:
it is powerful at extracting information from complex non-Gaussian fields, but it is deterministic, interpretable, and does not require any training. It compresses information into a relatively small set of coefficients due to its relation to wavelets, and it is a `first-order' statistic, i.e. it does not use higher powers of the input field, which makes the scattering coefficients robust statistics.

On the one hand, the scattering transform can be understood as a CNN with pre-determined, non-trainable kernels. It inherits CNN's way to extract spatial patterns, including local convolutions, a non-amplifying nonlinear operation, and multi-layers, but it abandons the learning ability which makes CNNs less controllable or interpretable.

On the other hand, as discussed in \citet{Cheng_2020}, the scattering transform is also similar to $N$-point functions averaged within wide, logarithmically-spaced bins, but with fundamental differences. Instead of using the multiplications of field intensities, the scattering transform uses a `low-order' nonlinear operation to convert field fluctuations into their strengths. Therefore, the scattering transform does not amplify the tail of the underlying distribution function and thus alleviates the corresponding information loss problem \citep{Carron_2011, Carron_2012, Carron_2012ApJ} of higher-order statistics. We will present a more in-depth comparison between the higher-order moments and scattering statistics in a forthcoming paper (Cheng \& M\'enard, in prep).

\subsection{Formulation}
\label{sec:formulation}

The mathematics of scattering transform was introduced in \cite{Mallat_2012, Bruna_2013}. Here, we present a succinct introduction, following the notation in our first paper \citep{Cheng_2020}. 
In brief, the scattering transform extracts information from a field by applying the following operations: 
\begin{itemize}
    \item wavelet convolution + modulus
    \item hierarchical structure
    \item spatial average
\end{itemize} 
It can be formalized in a recursive way, as illustrated in Figure~\ref{fig:ST_tree}:
\begin{align}
    I_0\,\,&\equiv \text{input field}\\
    I_n^{j,l} &\equiv \left|I_{n-1} \star \psi^{j,l}\right|\,\text{ (the scattering operation)}\label{eq:I_n}\\
    S_n &\equiv \langle I_n \rangle\,,
\end{align}
where $\psi^{j,l}$ stands for a wavelet indexed by its scale $j$ and orientation $l$. 

The scattering operation is composed of a wavelet convolution followed by a pixelwise modulus. The function of convolution $\star \psi^{j,l}$ is to separate fluctuations into different scales; the function of modulus $\left|\cdot\right|$ is to convert the selected fluctuations into their local strengths. Succesive applications of the scattering operation then form a tree structure, i.e. a planar multi-layer network, with various scattering fields $I_n$ at its nodes. Each $I_n$ is the intensity map of fluctuations of a particular scale in the previous-order field $I_{n-1}$. It can also be understood as a non-linear intensity map of a certain pattern in the input image $I_0$.

Finally, the translation-invariant scattering coefficients $S$ are defined as the spatial average of every scattering image, which resembles the pooling operation in convolutional neural networks. In this way, the 0th-, 1st-, and 2nd-order scattering coefficients can be written explicitly as:
\begin{align}
    S_0 &\equiv \langle I_0 \rangle \label{eq:S0}\\
    S_1^{j_1,l_1} &\equiv \langle I_1^{j_1,l_1}~~~~~~~\rangle = \langle \left|I_0\star\psi^{j_1,l_1}\right| \rangle \label{eq:S1}\\
    S_2^{j_1,l_1,j_2,l_2} &\equiv \langle I_2^{j_1,l_1,j_2,l_2}\rangle = \langle  \left| |I_0\star\psi^{j_1,l_1}|\star\psi^{j_2,l_2}\right| \rangle\,. \label{eq:S2}
\end{align}

The scattering operation which separates scales and measures fluctuation strength is qualitatively similar to the power spectrum analysis. Indeed, if we replace the modulus in Equation~\ref{eq:I_n} by modulus squared, the scattering coefficients $S_n$ will exactly become averaged $2^n$-point functions weighted by a sequence of wavelets. However, the modulus instead of modulus squared makes it much more efficient in extracting information from the tail of the field probability distribution function (PDF). 

The number of scattering coefficients $S_n$ is determined by the number of distinct wavelets used. Using $J$ scales and $L$ orientations, the total number of coefficients up to the 2nd order is $1 + JL + J(J-1)L^2/2$.
If orientation does not provide much relevant information, one may average over the directional indices:
\begin{align}
    s_n &\equiv \langle S_n\rangle _{l_1,...,l_n},
    \label{eq:s_n}
\end{align}
which reduces the number of coefficients to $1 + J + J(J-1)/2$, up to the 2nd order. For example, when using dyadic wavelets, a 512$\times$512 pixel image has 8 wavelet scales ($J$ = 8), corresponding to only 37 reduced scattering coefficients. 

For more intuitive understanding of the scattering transform and coefficients, we refer the readers to \citet{Cheng_2020}; For more mathematical properties, we refer the readers to \citet{Mallat_2012} and \citet{Bruna_2015}.

\subsection{Image generation based on scattering coefficients}
\label{sec:synthesis}

To compare the performance of different statistics in characterizing a random field, it is informative to compare them through their image synthesis capabilities. Here we will focus on three statistics: the power spectrum, the bispectrum and the scattering coefficients in the context of weak lensing convergence maps. The idea of synthesis is to randomly generate new images (or, in other words, to sample from the ensemble of images) that have the same summary statistics as the target one, and then visually check the \emph{texture} of the generated images \citep[see, e.g.,][]{bruna2019multiscale, Allys_2020}.

To implement it, we start from a random image and modify it in order to minimize the difference of the summary statistics between the generated and the target images. Technically, we start from a Gaussian random field that has the same power spectrum as the target image, and then use the `adam' optimizer in the python package \texttt{torch.optim} to minimize a loss function defined from the difference of summary statistics between the generated and target image. In the scattering coefficient and bispectrum cases, we also minimize the difference of their $L_1$ norms and set a lower bound for all pixels. Our code for image generation is available online\footnote{\href{https://github.com/SihaoCheng/scattering_transform}{https://github.com/SihaoCheng/scattering\_transform}}.

In Figure~\ref{fig:synthesis}, we show the generated images using power spectrum $P$, bispectrum $B$, and the scattering coefficients $S$. For each set of statistics, four realizations are shown to illustrate the sampling variance. Compared to the power spectrum and bispectrum, the results from scattering coefficients look much more similar to a real lensing convergence map (target image), especially in the textures created by halos on all scales.

There are several caveats about image synthesis to keep in mind. For example, the image quality somewhat depends on the initial condition and the choice of loss function. More importantly, the concept of information must always be related to a particular task. In the image synthesis case, this task is `to distinguish fields under the metric of human eyes and brains', which can be different from the Fisher information in the cosmological parameter inference task. Nevertheless, the striking visual similarity between the target and generated images using scattering coefficients is evidence of their power to characterize lensing map textures.

\section{Cosmological Forecast}
\label{sec:method}

To explore the constraining power of the scattering coefficients on cosmological parameters, we calculate the Fisher forecast \citep{Fisher_1935, Vogeley_1996, Tegmark_1997} and Bayesian posterior of the cosmological parameters, after building a likelihood emulator for the scattering coefficients using simulated lensing convergence maps.

\subsection{Simulated lensing maps}
\label{sec:data}

To explore the effects of dark energy and neutrino mass on lensing scattering coefficients, we use two sets of simulated convergence maps, which were generously made available by the Columbia lensing team\footnote{\href{http://columbialensing.org}{http://columbialensing.org}}. Both datasets were designed for probing the non-Gaussian information in weak lensing cosmology. The convergence maps were produced through ray-tracing $N$-body simulations to certain source redshifts without the Born approximation, using the \texttt{LensTools} python package \citep{Petri_2016}. All the cosmologies are spatially flat. 
The first dataset, described in \citet{Petri_2016, Matilla_2016, Liu_2016}, varies dark energy properties and matter density ($w_0$, $w_a$, and $\Omega_m$). The convergence is traced to one source redshift ($z$ = 2). 
The second dataset, \texttt{MassiveNuS} \citep{Liu_2018}, varies neutrino mass, matter density, and primordial fluctuation amplitude ($M_\nu$, $\Omega_m$, and $A_s$, equivalant to $M_\nu$, $\Omega_m$, and $\sigma_8$), and the convergence is traced to five source redshifts ($z$ = 0.5, 1, 1.5, 2, 2.5). Both datasets contain multiple realizations of convergence maps in multiple cosmologies, with 3.5$\times$3.5 deg$^2$ field of view (where non-linear gravitational evolution becomes important) and 512$\times$512 pixel resolution. Thus these maps cover multipole moments $l$ from 100 to roughly 30,000. Below we describe these datasets in more detail.

\textbf{Dark Energy simulations:} This dataset contains simulations of 7 $w_0w_a$CDM cosmologies with different dark energy properties, including the present dark energy density $\Omega_\Lambda$ and the dark energy equation of state index $w(a)$, parametrized by $w_0$ and $w_a$ through $w(a)=w_0+(1-a)w_a$ \citep{Chevallier_2001}. Each simulation was run in a 240~Mpc/$h$ box with 512$^3$ particles and used to generate 1,024 mock convergence maps with 3.5$\times$3.5 deg$^2$ field of view. The lensing sources were set at redshift $z$ = 2. We down-sampled the original 2048$^2$ pixel maps to a 512$^2$ resolution with 0.41 arcmin per pixel, by averaging adjacent pixels. Besides the dark energy properties, other cosmological parameters are fixed: baryon density $\Omega_b$~=~0.046, Hubble constant $h$~=~0.72, scalar spectral index $n_s$ = 0.96, normalization of fluctuation amplitude $\sigma_8$~=~0.8, effective number of relativistic degrees of freedom $n_\text{eff}$~=~3.04, total neutrino masses $M_\nu$ = 0.0, and temperature of the cosmic microwave background $T_\text{CMB}$~=~2.725~K. The dark matter density is set so that the universe is spatially flat, i.e., $\Omega_m$ = $1-\Omega_\Lambda$. Pipelines for the $N$-body simulations and ray-tracing are described in \citet{Petri_2016}, \citet{Matilla_2016}, and \citet{Liu_2016}. Specifications of the simulations are also listed on the website of the Columbia Lensing team. 

\textbf{Neutrino mass simulations (\texttt{MassiveNuS}):} Generated by \citet{Liu_2018}, this dataset contains convergence maps for several source redshifts created from 101 $\Lambda\nu$CDM simulations with different $M_\nu$, $\Omega_m$, and $A_s$ (equivalent to $M_\nu$, $\Omega_m$, and $\sigma_8$). Other cosmological parameters are fixed: $\Omega_b$~=~0.046, $h$~=~0.70, $n_s$~=~0.97, $w$~=~--1. Each simulation was run in a 512~Mpc/$h$ box with 1024$^3$ particles and used to generate 10,000 convergence maps with 3.5$\times$3.5 deg$^2$ field of view and 512$^2$-pixel resolution. The fiducial cosmology has parameters ($M_\nu, \Omega_m, A_s$) = (0.1 eV, 0.3, 2.1$\times$10$^{-9}$). One cosmology with massless neutrinos ($M_\nu, \Omega_m, A_s$) = (0 eV, 0.3, 2.1$\times$10$^{-9}$) is used to calculate the covariance matrix of summary statistics. In this study, we will use the convergence maps for 5 source redshifts ($z$ = 0.5, 1.0, 1.5, 2.0, and 2.5), in 28 cosmologies that are close to the fiducial cosmology, i.e. in the range of $A_s\in[1.8,2.7]\times 10^{-9}$, $M_\nu \in[0.06,0.60]$ eV, $\Omega_m\in[0.28,0.32]$. This range of parameters is enough for estimating the scattering coefficients' constraining power for cosmological parameters in a Rubin-observatory-like survey.

In real lensing surveys, the galaxy shape noise will dilute cosmological information on small scales, acting as a scale cut. For simplicity, we approximate such noise by Gaussian white noise on convergence maps \citep{vanWaebeke_2000}, $\sigma_\text{noise}^2=\langle \sigma_\epsilon ^2 \rangle / n_\text{gal} A_\text{pix}$, where $A_\text{pix}$ is the pixel area. We adopt an averaged ellipticity squared $\langle \sigma_\epsilon ^2 \rangle = 0.3^2$ and a number density of background galaxies $n_\text{gal}$ of 44 per arcmin$^2$ to simulate the weak lensing survey of the Rubin Observatory \citep{LSST_2009}.
The multiple source redshifts of the \texttt{MassiveNuS} convergence maps also allow us to explore the information gain from redshift tomography. We consider two situations: (1) only the convergence maps for source redshift $z$ = 1 are used, with $n_\text{gal}$ = 44 per arcmin$^2$; (2) convergence maps for five source redshifts are used, with $n_\text{gal}$ = 8.83, 13.25, 11.15, 7.36, and 4.26 per arcmin$^2$, respectively. The source densities are derived from the expected source redshift distribution for the Rubin Observatory survey: $n(z) \propto z^2 \exp(-2z)$ with a normalization of 50 per arcmin$^2$ \citep{LSST_2009}. We calculate the galaxy number density in each fiducial redshift bins by integrating the distribution from z = 0.25 to 2.75 with steps of $\Delta z = 0.5$ as the bin widths. For noisy maps, we also carry out a smoothing with Gaussian kernel $\sigma$~=~1~arcmin. 
Finally, the effects of masks and imperfection in shear-to-convergence conversion \citep[e.g.,][]{Kaiser_1993, Pires_2020, Jeffrey_2020}, which are important practical problems for weak lensing cosmology, are not taken into account in this analysis. Our primary goal is only to compare the statistical properties of different estimators.

\begin{figure}
    \centering
    \includegraphics[width=\columnwidth]{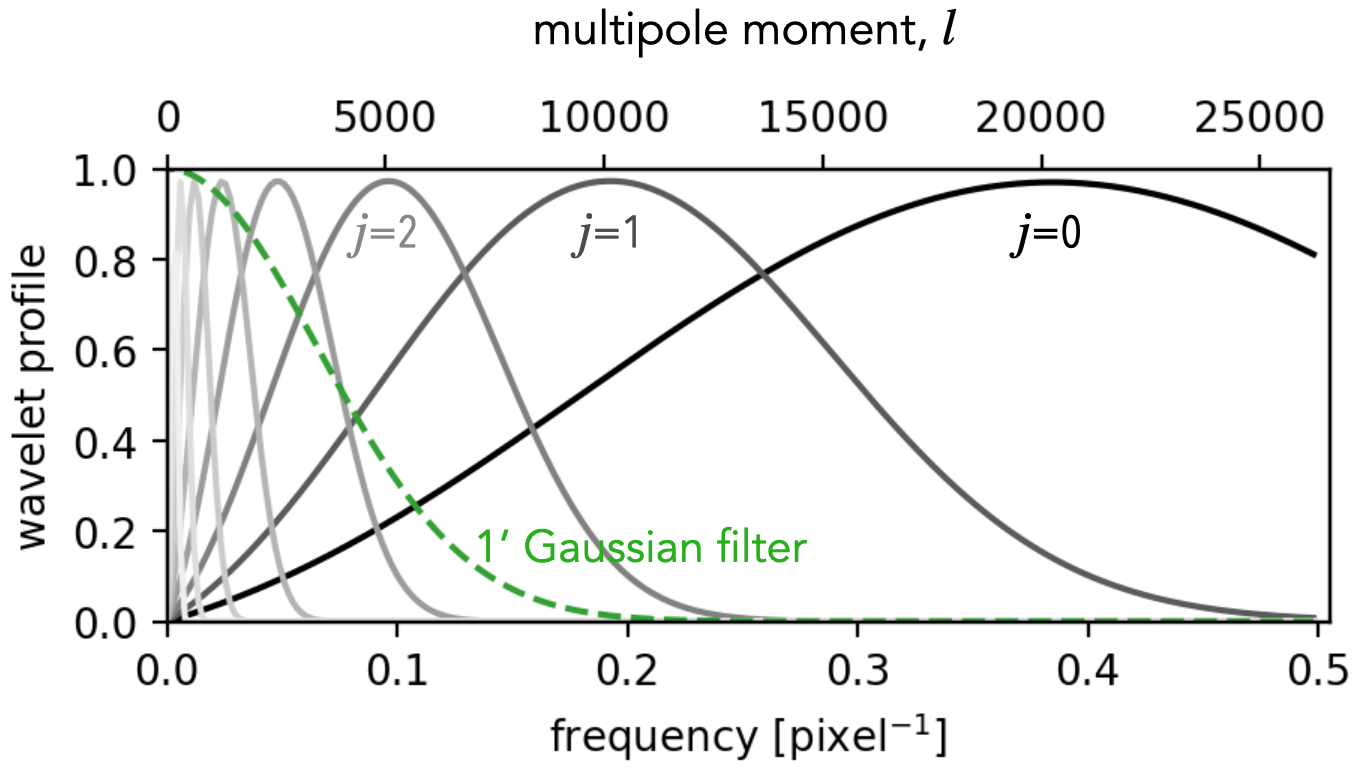}
    \caption{Scale coverage of wavelets used by the scattering transform. We also show the profile of a 1 arcmin Gaussian filter in frequency space for comparison.}
    \label{fig:scales}
\end{figure}

\subsection{Summary statistics}
\label{sec:statistics}

We calculate three sets of summary statistics from the simulated lensing convergence maps:
\begin{itemize}
    \item 30 binned power spectrum P($l$) which we sample with linear bins of scale,
    \item 125 binned bispectrum B($l_1$,~$l_2$,~$l_3$) which are sampled within 10 linear bins for each of $l_1$, $l_2$, and $l_3$ in the bispectrum monopole B($l_1$, $l_2$, $l_3$), and
    \item 36 reduced scattering coefficients $s_1^{j}$ and $s_2^{j_1, j_2}$.
\end{itemize}
For the scattering coefficients, we use the same Morlet wavelets as described in \citet{Cheng_2020}, with 4 azimuthal orientations ($L$~=~4) and 8 dyadic scales ($J$~=~8). The scales correspond to 8 logarithmic bins from $l$~=~100 to 37,000, as shown in Figure~\ref{fig:scales}. Following Equation~\ref{eq:s_n}, we average over the azimuthal orientations and use the 36 reduced scattering coefficients $s_1$ and $s_2$. On noisy maps, the small-$j$ (small scale, high-$l$) scattering coefficients are dominated by noise, and we verify that abandoning the small-$j$ coefficients does not change the precision of cosmological inference.
As a result, the set of informative scattering coefficients on noisy maps is even more compact than on noiseless maps, which is desirable for inference analysis. As these scattering coefficients are non-negative, we instead consider their logarithm. This step helps to Gaussianize their PDF, which is a desired property for likelihood inference. We have also checked that this transformation has negligible effects on cosmological parameter inference. In the rest of the paper, we will always consider the logarithm of the coefficients but simply refer to them as scattering coefficients for brevity. We point out that the log-transform cannot be applied to the bispectrum coefficients which can be both positive and negative.


In summary, for each 3.5$\times$3.5 deg$^2$ lensing map, we calculate 36 scattering coefficients, 30 binned power spectrum coefficients, and 125 binned bispectrum coefficients for the cosmological parameter inference.
We have developed a python3 package optimized in speed to calculate the translation-invariant scattering coefficients. The package, together with codes calculating the bispectrum, are available online\footnote{\href{https://github.com/SihaoCheng/scattering_transform}{https://github.com/SihaoCheng/scattering\_transform}}. We estimated the power spectrum using the \texttt{LensTools} python package \citep{Petri_2016}.

\subsection{Likelihood and parameter inference}
\label{sec:likelihood}

Parameter inference requires a likelihood function $p(\bm{x}|\bm{\theta})$. We first assume that its function form is Gaussian, i.e., the probability distribution function (PDF) of the summary statistics $\bm{x}$ is a multivariate Gaussian distribution for a given cosmology $\bm{\theta}$. We also assume the cosmological dependence of the mean vector of the Gaussian PDF can be approximated by a smooth function in the range of interest. For the dark energy dataset, we use a linear dependence, because there are only 7 cosmologies. For the \texttt{MassiveNuS} dataset, we use second-order polynomials. We set the covariance matrix to be independent of cosmology. The latter assumption is conservative but robust when the likelihood is not exactly Gaussian \citep{Carron_2013}. The cosmological dependence of mean vector and the covariance matrix are estimated using the simulated convergence maps. 

Given the likelihood, we use the Fisher matrix and Bayesian posterior to quantify the constraining power of lensing scattering coefficients for cosmological parameters. The mathematical formulations of the inference frameworks are given in Appendix~\ref{app:inference}.
The Fisher matrix allows us to forecast the constraining power using only on the local cosmological dependence of the likelihood. For the noiseless cases we consider in this paper, it provides almost identical results as the posterior. However, with increasing noise level, the difference may become non-negligible. We will therefore use the Bayesian posterior for the results of noisy data.

\begin{figure}
    \centering
    \includegraphics[width=\columnwidth]{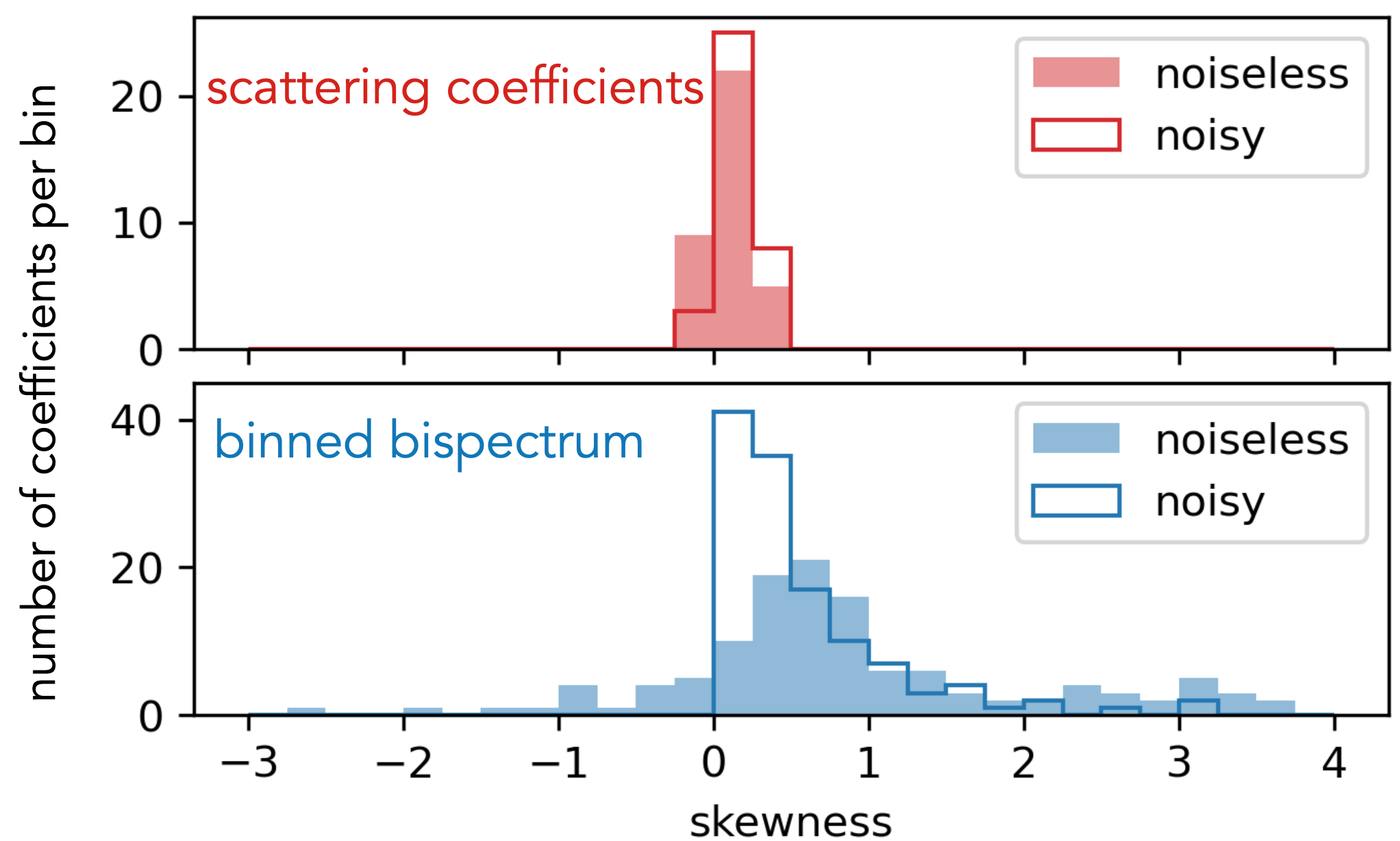}
    \caption{
    Skewness distribution of the (log) scattering and (binned) bispectrum coefficients, measured from 10,000 realizations of convergence maps at the fiducial cosmology of MassiveNuS simulation set. If the PDF of these statistical coefficients are Gaussian, then the skewness should be zero.
    }
    \label{fig:PDF}
\end{figure}

\begin{table}
 \caption{\label{tab:PDF}Number of directions that are significantly non-Gaussian in the high-dimension PDF of summary statistics. Values outside and inside parentheses are for noiseless and noisy maps, respectively. Rows are different criteria to identify non-Gaussianity, which systematically suggest that the scattering coefficients have nearly-Gaussian PDF, whereas the bispectrum coefficients do not.}
 \begin{tabular}{c|cc}
  \hline
    summary statistics & log scattering coefficients & binned bispectrum\\
    number of coefficients & 36 & 125\\
  \hline
  \hline
    |skewness| $>$ 0.5 & 0 (0) & 87 (49)\\
    |skewness| $>$ 1 & 0 (0) & 45 (22)\\
    kurtosis $>$ 0.5 & 1 (1) & 120 (78)\\
    kurtosis $>$ 1 & 0 (0) & 111 (49)\\
  \hline
    whitened, |skewness| $>$ 0.5 & 0 (0) & 18 (4)\\
    whitened, |skewness| $>$ 1 & 0 (0) & 9 (2)\\
    whitened, kurtosis $>$ 0.5 & 0 (0) & 124 (25)\\
    whitened, kurtosis $>$ 1 & 0 (0) & 119 (11)\\
  \hline
 \end{tabular}
\end{table}

\begin{figure*}
    \centering
    \includegraphics[width=1.09\columnwidth]{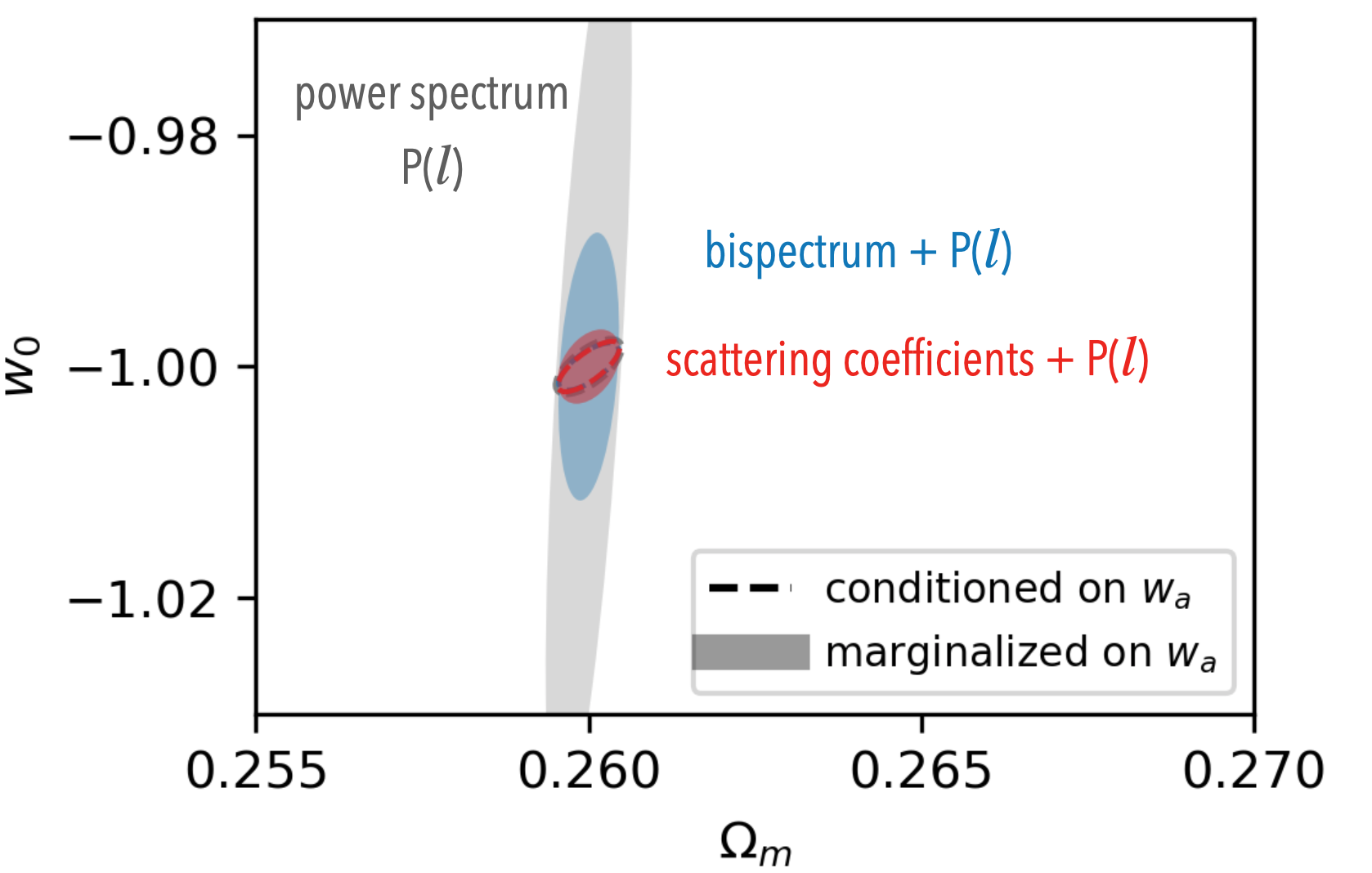}
    \includegraphics[width=0.97\columnwidth]{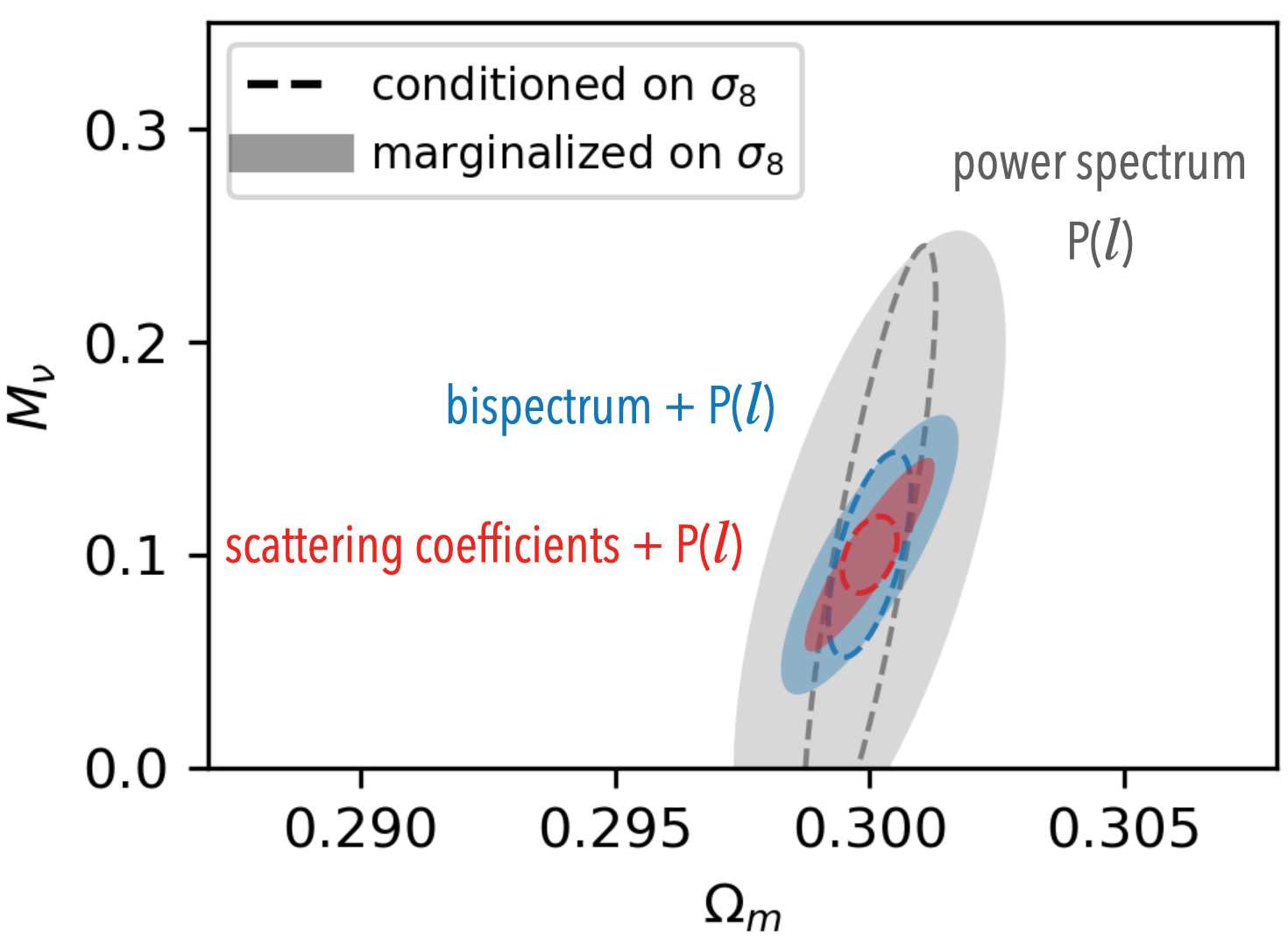}
    \caption{Fisher forecast with noiseless $\kappa$ maps (95\% confidence ellipses). The scattering coefficients have remarkably high constraining power for cosmological parameters. \emph{Left:} Forecast for dark energy equation of state index $w(a) = w_0 + (1-a)w_a$, from 20,000 deg$^2$ noiseless map with $l$ nominally up to roughly 30,000 and source redshift $z$ = 2. \emph{Right:} Forecast for neutrino mass $M_\nu$, from 20,000 deg$^2$ noiseless map with source redshift $z$ = 1.}
    \label{fig:noiseless}
\end{figure*}

\section{Results}
\label{sec:results}

\subsection{The sampling distribution of summary statistics}
\label{sec:PDF}

Before exploring the constraining power on cosmological parameters, we first present one key advantage for using the scattering coefficients, namely that their probability density distribution (PDF) or sampling distribution is well Gaussianized.

For many cosmological studies, the typical inference framework assumes 
a Gaussian PDF for the statistical estimator. This is the so-called Gaussian likelihood assumption, though strictly speaking, likelihood is $p(\bm{x} | \bm{\theta})$ with the data $\bm{x}$ fixed, while in practice one usually parametrizes and fits the PDF (the same expression with the model parameters $\bm{\theta}$ fixed) to simulations. So we will call it the Gaussian PDF assumption. This assumption comes from the central limit theorem when the field of view tends to infinity. However, with a finite field of view, there is no guarantee for a Gaussian PDF \citep[e.g.,][]{Sellentin_2018, Hahn_2019likelihood, DiazRivero_2020, Jeffrey_2021}, and approximating a non-Gaussian PDF with a Gaussian one may introduce bias and/or underestimation of uncertainty to parameter inference. Therefore, summary statistics that Gaussianize quickly are favoured.
Note that the non-Gaussianity of the field is not the same as the non-Gaussianity of the summary statistics. Some statistics of a highly non-Gaussian field can be nearly Gaussian distributed due to central limit theorem. On the other hand, statistics of a Gaussian random field can have non-Gaussian sampling distribution. For example, the power of each Fourier mode is expected to have a $\chi^2$-distribution.

Deviations from the Gaussian PDF approximation are harmful to inference, among which the most problematic is the presence of a heavy tail. To quantify this effect, we will consider both the skewness and kurtosis parameters of the 1D marginal distributions. More precisely: we calculate these parameters for both noiseless and noisy maps, using the fiducial cosmology of \texttt{MassiveNuS} simulation with source redshift $z$ = 1. The skewness is defined as $\langle(\frac{x - \mu_{x}}{\sigma_{x}})^3 \rangle$ for each summary statistic $x$, where $\mu$ and $\sigma$ are the sample mean and standard deviation calculated from the 10,000 realizations. It measures the asymmetry and tail heaviness of the one-variable PDFs, which is one possible deviation from a Gaussian distribution. The scattering coefficients\footnote{
We point out that, even without taking the logarithm of the scattering coefficients, the skewness values of the 'raw' scattering coefficients are still relatively small, typically between $0$ and $1$.} have much lower skewness than the bispectrum coefficients. We show the histograms of their skewness parameters in Figure~\ref{fig:PDF}. Similarly, we measure the kurtosis defined as $\langle (\frac{x - \mu_{x}}{\sigma_{x}})^4\rangle - 3$, which is more sensitive to the tail heaviness. For a 1D Gaussian distribution, given our sample size, the skewness and kurtosis should be 0 $\pm$ 0.03 and 0 $\pm$ 0.05, respectively. The measured values for the scattering coefficients are indeed much closer to these expectations than the values obtained from the bispectrum which, in some cases, exceed the upper bound by 
more than two orders of magnitude. The effect of global whitening (using principle component analysis to remove correlations between dimensions) is also examined. Under all these metrics, the scattering coefficients show a low non-Gaussianity. In contrast, many dimensions of the bispectrum data vector are asymmetric or heavy tailed, even on the noisy maps where the field itself is more Gaussian. This is shown in Table~\ref{tab:PDF} where we present the number of coefficients exceeding a given skewness or kurtosis threshold for each estimator. Tests using the 95-percentile, yet another measure of tail heaviness, show similar results. Assessing the non-Gaussianity of parameters in high-dimensional spaces can be done from numerous points of view. For example, \citet{Sellentin_2018} proposed to use all pairs of 2D marginal PDFs. Such tests will be valuable in future studies aimed at inferring cosmological parameters.

The difference in the types of distributions found for scattering and bispectrum coefficients can be understood from their respective designs. The bispectrum coefficients are products of three random variables. The multiplication in general yields a distribution with heavier tail than the original variables. In contrast, the scattering transform uses a `first-order' modulus which does not amplify the tail. Therefore, as the field of view increases, the central limit theorem Gaussianizes the scattering coefficients much quicker than bispectrum coefficients. Moreover, the logarithmic binning of scales through wavelets combines more Fourier modes, which also helps the Gaussianization process.

In summary, we find that the scattering coefficients Gaussianize particularly well, which is desirable and necessary for accurate cosmological inference.

\subsection{Constraining cosmological parameters}
Having checked the Gaussian PDF assumption for the scattering coefficients, now we compare the cosmological constraints obtained by different summary statistics. These constraints were obtained based on simulated convergence maps with varied cosmological parameters, and 1,000 (for the dark energy simulation set) or 10,000 (for the neutrino mass simulation set \texttt{MassiveNuS}) realizations for each cosmology. 

Because the convergence maps have a long-tailed PDF (approximately log-normal), we do expect the scattering coefficients to extract more information than the bispectrum. Indeed, we find a significantly tighter constraint using the scattering coefficients on noiseless maps (when all high-$l$ modes are included).
In Figure~\ref{fig:noiseless} we show the Fisher forecast of cosmological parameters using noiseless convergence maps, which corresponds to $l$ from 100 to roughly 30,000. For simplicity, we only show a plane of two parameters for each dataset. In both panels of Figure~\ref{fig:noiseless}, the marginalized ellipses are the projection of the three-parameter Fisher ellipsoids onto the plane, while the conditioned ellipses are their cross-sections. 

For dark energy parameters (left panel), the power spectrum constrains $w_0$ well when $w_a$ is known (conditioned), but constrains poorly if $w_a$ is also to be constrained (marginalized), meaning a strong degeneracy between $w_0$ and $w_a$. Compared to the bispectrum, which is known to partly break this degeneracy \citep{Takada_2003}, the scattering coefficients almost eliminate it in the noiseless case.
In weak lensing cosmology, there is another well-known degeneracy between $\Omega_m$ and $\sigma_8$, which the scattering coefficients can also break \citep{Cheng_2020}. However, the dark energy dataset used in this study has a fixed $\sigma_8$ value. So, it remains to be investigated whether or not the scattering coefficients can break both degeneracies at the same time.

On the noiseless maps, scattering coefficients also provide much tighter constraint on the neutrino mass sum $M_\nu$ than the power spectrum and bispectrum (see the right panel of Figure~\ref{fig:noiseless}). In the parameter space spanned by $\Omega_m$, $\sigma_8$, $M_\nu$, the power spectrum's constraining power comes mainly from its overall amplitude, which results in a disc-like Fisher ellipsoid. Adding the bispectrum reduces the disc's area by partly breaking the $\Omega_m$--$\sigma_8$ degeneracy, whereas adding the scattering coefficients almost eliminate it, reducing it to a one-dimension instead of two-dimension degeneracy. This is consistent with the finding presented in the figure 5 of \cite{Cheng_2020}. The ratio between the volumes of the three-parameter Fisher ellipsoids (i.e., the square root of determinant of the covariance matrix) shown in the right panel of Figure~\ref{fig:noiseless} is roughly 20:4:1.

\begin{figure}
    \centering
    \includegraphics[width=0.99\columnwidth]{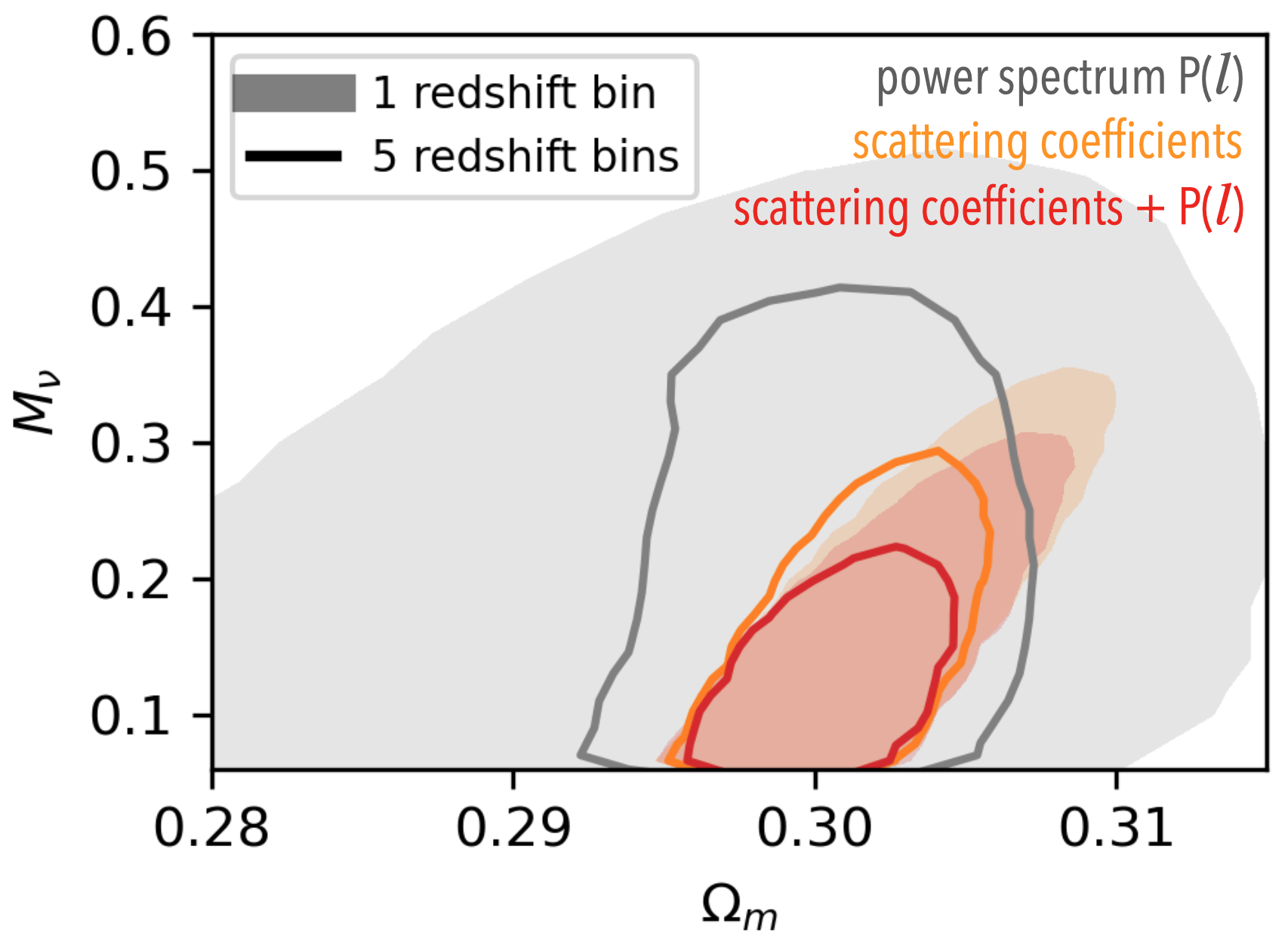}
    \caption{Forecast of cosmological parameters (95\% confidence contours) for a Rubin-observatory-like survey with 20,000 deg$^2$ field of view, with $M_\nu$, $\Omega_m$, and $A_s$ (or $\sigma_8$) to be constrained. Colours represent results of different sets of summary statistics of lensing convergence ($\kappa$). Improvement from adding redshift tomography is also shown.}
    \label{fig:1z5z}
\end{figure}

When the galaxy shape noise is taken into account, the constraining power of all summary statistics weakens due to the loss of small scale (high-$l$) information.
In Figure~\ref{fig:1z5z} we show the forecast of neutrino mass sum ($M_\nu$) for a Rubin-observatory-like survey, using the Bayesian posterior with fiducial cosmology at ($M_\nu$, $\Omega_m$, $A_s$) = (0.1 eV, 0.3, 2.1$\times$10${^{-9}}$). In spite of noises, there are still substantial non-Gaussian structures in the lensing convergence maps, thus the scattering coefficients still set much tighter cosmological constraints (more than two times in $M_\nu$) than using the power spectrum alone. 
In this noisy case, the combination of bispectrum and power spectrum sets a similar constraint to the combination of scattering coefficients and the power spectrum (though not shown in the figure). 
Figure~\ref{fig:1z5z} also show the effect of redshift tomography. A five-bin redshift tomography improves the scattering coefficients' constraint for $M_\nu$ by about 40\%.

With the same dataset (\texttt{MassiveNuS}), similar explorations were performed for other non-Gaussian statistics, including the PDF \citep{Liu_2019, Boyle_2021}, bispectrum \citep{Coulton_2019}, peak count \citep{Li_2019}, starlet peak count \citep{Ajani_2020}, minima count \citep{Coulton_2020}, Minkowski functionals \citep{Marques_2019}, and starlet $l_1$-norm \citep{Ajani_2021}. They all found similar improvement from redshift tomography for their respective summary statistics.

\section{Conclusions}
\label{sec:conclusions}

In cosmology, substantial information is stored in non-Gaussian structures on small scales, which requires statistics beyond the power spectrum to extract. 
Motivated by ongoing and upcoming deep surveys, we explore the weak lensing application of a novel and powerful non-Gaussian statistics, called the scattering transform.
We extend the constraining forecast of $\Omega_m$ and $\sigma_8$ in \citet{Cheng_2020} to more cosmological parameters including the dark energy parameters $w_0$, $w_a$ and neutrino mass $M_\nu$.
To do so, we use mock convergence maps from two publicly available simulation suites (the dark energy and \texttt{MassiveNuS} sets) made by the Columbia Lensing team.

We first show that using the scattering coefficients, one can generate random images with textures very close to a simulated lensing map, which cannot be achieved by using the power spectrum and bispectrum coefficients.
Then, we show that the scattering coefficients provide significantly better constraint for dark energy parameters $w_0$, $w_a$ and neutrino mass $M_\nu$ than using power spectrum alone.

For noiseless maps (when high-$l$ modes are accessible), the scattering transform also outperforms the bispectrum and power spectrum. This result can be explained by the `first-order' nature of scattering coefficients which stay proportional to the field intensity, in contrast to higher-order statistics which amplify the distribution tail. This \emph{lower-order} nature together with the wavelet weighting strategy efficiently concentrates cosmological information into a compact set of scattering coefficients.
Moreover, the \emph{lower-order} nature makes the distribution of scattering coefficients much more Gaussian and robust than higher-order statistics, which is essential for accurate likelihood parametrization and parameter inference.

We also provide a forecast of $M_\nu$ with a noise level of the Rubin observatory survey, where the scattering coefficients set a 2 times tighter constraint than using the power spectrum alone, and redshift tomography improves the constraint by an additional 40\%. Similar forecast for the dark energy parameters is left for future study, due to the redshift limitation of the mock convergence maps.
For noisy maps, although the constraining power of the scattering coefficients is not significantly better than the combination of bispectrum and power spectrum, we argue that the scattering coefficients are still preferred, because the PDF of scattering coefficients is much more Gaussian than bispectrum coefficients.

The scattering transform yields a compact set of \emph{lower-order} summary statistics that are stable, powerful, and efficient for characterizing non-Gaussian structures. Together with \citet{Cheng_2020}, we have shown that the scattering transform has great potential to serve as the non-Gaussian summary statistics in observational cosmology.

\section*{Acknowledgements}
We thank the Columbia Lensing group (\href{http://columbialensing.org}{\url{http://columbialensing.org}})
for making their suite of mock lensing maps publicly available and NSF for supporting the creation of those maps through grant AST-1210877 and XSEDE allocation AST-140041. We thank Yuan-Sen Ting for his help on image synthesis and Douglas Finkbeiner for pointing out an error in the first version of this paper. S. C. thanks Siyu Yao for her constant inspiration and encouragement.

\section*{Data availability}

The data underlying this article were accessed from the Columbia Lensing group (\href{http://columbialensing.org}{\url{http://columbialensing.org}}). The derived data generated in this research will be shared on reasonable request to the corresponding author.

\bibliographystyle{mnras}
\bibliography{ST_DE}

\appendix

\section{Binned bispectrum}
\label{app:bispectrum}

Here we describe the formulation of the binned 
bispectrum.
%
The bispectrum is the products of three Fourier coefficients,
\begin{align}
    B(\bm{k}_1, \bm{k}_2, \bm{k}_3) \equiv \langle \tilde{I}(\bm{k}_1) \tilde{I}(\bm{k}_2) \tilde{I}(\bm{k}_3) \rangle
\end{align}
Because of symmetry, it is enough to consider only cases with $k_1 < k_2 < k_3$.
Under statistical homogeneity, the coefficients vanish unless $\bm{k}_1+\bm{k}_2+\bm{k}_3=0$.
To reduce the number of coefficients, one can use the binned bispectrum mono-pole, which averages the bispectrum according to the magnitude of $k_1$, $k_2$, and $k_3$ 
\begin{align}
    B_{i,j,k} =& \int_{k_{i}}^{k_{i+1}}d\bm{k}_1 \int_{k_{j}}^{k_{j+1}}d\bm{k}_2 \int_{k_{k}}^{k_{k+1}}d\bm{k}_3 \nonumber\\ &(2\pi)^2\delta^{(2)}(\bm{k}_1+\bm{k}_2+\bm{k}_3) \cdot \tilde{I}(\bm{k}_1) \tilde{I}(\bm{k}_2) \tilde{I}(\bm{k}_3)
    \label{eq:B}
\end{align}

It is not convenient to select $\delta(\bm{k}_1+\bm{k}_2+\bm{k}_3)$ triplets within each bin. Fortunately, expanding the right-hand side of Eq.~\ref{eq:B} in real space will simplify the integrant,
\begin{align}
    B_{i,j,k} = \langle (I\star f_i) (I\star f_j) (I\star f_k) \rangle\,,
\end{align}
where $f_i(\bm{x}) = \int_{k_i}^{k_{i+1}} e^{i\bm{k}\cdot\bm{x}} d\bm{k}$ is the inverse Fourier transform of the binning, an annulus in Fourier space and a wavelet-like profile in real space.
This form of expression, similar to the scattering coefficients, means that the bispectrum (and other $N$-point functions) can also be expressed as the spatial mean of some non-linear transformation of the input field, though it has different properties from the scattering one.



\section{Cosmological inference framework}
\label{app:inference}

For a statistical model of observable $\bm{x}$, the Cram\'er--Rao inequality sets the lower limit of any unbiased estimators of model parameters obtained from the observable, therefore it can be used to quantify the uncertainty of parameter inference.
\begin{align}
    \text{cov}(\hat{\bm{\theta}}) \geq \mathbfss{I}(\bm{\theta})^{-1}\,,
\end{align}
where $\mathbfss{I}(\bm{\theta})$ is the Fisher information matrix, whose elements are defined as
\begin{align}
    \I_{m,n}(\bm{\theta}) \equiv -\left\langle \frac{\partial \text{ln}\,p(\bm{x}|\bm{\theta} )}{\partial \theta_m} \frac{\partial \text{ln}\,p(\bm{x}|\bm{\theta} )}{\partial \theta_n}\right\rangle\,.
\end{align}
In our case, $\bm{\theta}$ is the cosmological parameters, and $\bm{x}$ is the vector of summary statistics such as the scattering coefficients. The PDF is assumed to be Gaussian,
\begin{align}
\label{eq:Gaussian_likelihood}
    p(\bm{x}|\bm{\theta} ) \propto \frac{1}{\sqrt{|\mathbfss{C}|}}\text{exp}[-\frac12 (\bm{x}-\bm{\mu})^T\mathbfss{C}^{-1}(\bm{x}-\bm{\mu})]\,,
\end{align}
where $\mathbfss{C}(\bm{\theta})$ is the covariance matrix of summary statistics. We further assume $\mathbfss{C}(\bm{\theta})$ is cosmology independent. Thus, the Fisher matrix becomes
\begin{align}
    I_{m,n} =& \frac{\partial \bm{\mu}^T}{\partial \theta_m} \mathbfss{C}^{-1} \frac{\partial \bm{\mu}}{\partial \theta_n} \,.\label{eq:Fisher_Gaussian}
\end{align}
We use linear or second-order polynomials to fit the cosmological dependence of the mean vector $\bm{\mu}(\bm{\theta})$, and use the sample covariance matrix $\widehat{\mathbfss{C}}$ estimated from simulations at fiducial cosmologies ($\Omega_m$ = 0.26, $w_0$ = --1, $w_a$ = 0 for the dark energy dataset and $\Omega_m$ = 0.3, $M_\nu$ = 0, $A_s$ = 2.1$\times$10$^{-9}$ for the \texttt{MassiveNuS} dataset) to estimate the inverse of covariance matrix \citep{Hartlap_2007}: 
\begin{align}
    \widehat{\mathbfss{C}^{-1}} = \frac{N-D-2}{N-1}\widehat{\mathbfss{C}}^{-1}\,,
\end{align}
where $\widehat{\mathbfss{C}^{-1}}$ is an unbiased estimator, $N$ is the number of independent sample used for the estimation (1,000 for the dark energy dataset and 10,000 for the \texttt{MassiveNuS} dataset), $D$ is the number of summary statistics used, which is 36, 30, and 125 per redshift bin, for scattering coefficients, binned power spectrum, and binned bispectrum, respectively. The covariance matrix $\mathbfss{C}$ of summary statistics is measured from 3.5$\times$3.5 deg$^2$ maps. To make a forecast for Rubin-like surveys with 20,000 deg$^2$, we simply scale the covariance matrix by a factor of 3.5$\times$3.5/20,000, which is equivalent to assuming that each 3.5$\times$3.5 deg$^2$ patches are independent. Though not exactly correct, this is still a common approximation to get a quick forecast without running many full-sky simulations \citep[e.g.,][]{Li_2019, Liu_2019, Coulton_2019}.

The Bayesian posterior is another way to quantify the uncertainty of model parameter inference. Given the likelihood $p(\bm{x}|\bm{\theta})$, the posterior probability distribution of model parameter is
\begin{align}
    p(\bm{\theta}|\bm{x}) \propto p(\bm{x}|\bm{\theta}) p(\bm{\theta})\,,
\end{align}
where $p(\bm{\theta})$ is the prior distribution of $\bm{\theta}$. For the noisy cases with \texttt{MassiveNuS}, we use a flat prior in $\Omega_m\in[0.28,0.32]$, $M_\nu\in[0.06, 0.6]$ eV, and $A_s\in[1.8,2.7]\times10^{-9}$ to avoid extrapolation. The posterior is then sampled using \texttt{emcee}  \citep{Foreman-Mackey_2013}, a Markov Chain Monte Carlo (MCMC) sampler. We used 32 walkers, ran 1,000 steps to burn-in and then 10,000 steps to sample the posterior. Sampling convergence was checked within and among chains. For all combinations of summary statistics used in Figure~\ref{fig:1z5z}, the auto correlations of the chain are measured to be within $\pm$ 2 \% when the distance is longer than 100 steps.

\bsp	
\label{lastpage}
\end{document}